\documentclass[12pt,a4paper]{article}
\usepackage{epsfig}
\title{Liquid Drop Model with Different Neutron versus Proton Deformations
\thanks{The work was partially sponsored by the Polish Committee of Scientific
Research KBN No.~2P~03B~115~19, POLONIUM No.~017~04~UG (2001)}}

\author{    A. Dobrowolski$^1$, K. Pomorski$^{1,2}$, J. Bartel$^2$
\\
{\it $^1$Katedra Fizyki Teoretycznej, Uniwersytet M.C. Sk\l odowskiej,
                               Lublin, Poland}
\\
      {\it $^2$IReS -- IN$_2$P$_3$ -- CNRS and Universit\'e Louis Pasteur,
                             Strasbourg, France}}                               
 
\begin{document}
 
\maketitle   

\noindent
\begin{abstract}
The nuclear binding energies for 28 nuclei including several isotopic chains 
with masses ranging from $A \!=\! 64$ to $A \!=\! 226$ were evaluated using 
the Skyrme effective nucleon-nucleon interaction and the Extended Thomas-Fermi 
approximation.
The neutron and proton density distributions are assumed in the form of Fermi
functions the parameters of which are determined so as to minimize the total 
binding energy of any given nucleus. The present study is restricted to 
quadrupole shapes, but the neutron and proton density distributions are 
free to have different deformations. A simple expression for the variation of 
the nuclear energy with the neutron--proton deformation difference is derived.
\end{abstract}  

\bigskip
\noindent
PACS number:21.60.Jz,21.10.Dr,21.10.-k,21.10.Pc 
                                                             \\[2.0ex]

\section{Introduction}

The analysis of the experimental data on electron and $\alpha$--particle 
scattering, pionic atoms, and annihilation of antiprotons show that, in most 
nuclei, neutrons and protons have slightly different r.m.s. radii \cite{Ba89}.
The main reasons for this difference are the Coulomb repulsion between
protons and unequal numbers of neutrons and protons.

Fully microscopic approaches of the Hartree-Fock (HF) type using the effective 
nucleon-nucleon interactions of the Skyrme or Gogny type \cite{Va72,DG80},
as well as the relativistic mean field theory \cite{Ba95} reproduce in a 
rather satisfactory way the experimental proton and neutron r.m.s. radii and 
their isotopic shifts \cite{Ba95,Po97,Wa98}. 

In deformed nuclei, neutron and proton density distributions are expected to 
have not only different radii, but also exhibit different shapes, i.e.
different quadrupole and higher multipole deformations. An analysis of
theoretical densities based on a surface multipole moment expansion shows that
significant differences between neutron and proton deformations often occur
both in the ground state \cite{Po97} as well as along the whole path
to fission \cite{Be00}. The Hartree-Fock-Bogolubov (HFB) calculation performed 
with the Gogny effective force in Ref.~\cite{Be00} for $^{232}$Th, $^{236}$U,
$^{238}$U and $^{240}$Pu have shown that the multipole deformations of
the proton and neutron density distributions of fissioning nuclei are far from
being equal. The relative difference between them exceeds often 10~\% and
undergoes large variations which means that the thickness of the neutron skin
does not remain constant as the fissioning nucleus elongates. The effect on the
nuclear binding energy of these deformation differences was found to reach 
approximately 1.5 MeV, with fluctuations of the order of 1 MeV \cite{Be00}. 
These variations are clearly not negligible compared to typical fission 
barrier heights. 

Until now a large majority of macroscopic-microscopic type calculations of 
potential energy 
surfaces of fissioning nuclei assume equal deformations of proton and neutron 
distributions for both densities and single-particle potentials
(see e.g. \cite{Sm95,Mo95,St99}). In view of the results 
obtained in  Ref.\cite{Be00}, one can expect that such calculations predict 
fission barriers that are systematically about 1 MeV too high. 
In a sens things are even worth in the case of macroscopic-microscopic 
calculations as compared to the selfconsistent Hartree-Fock approach where 
deformation energy surfaces are usually generated through calculations with 
a constraint on some multipole moments of the mass distribution, which leaves 
protons and neutrons free to deform differently within this constraint, even 
though it of course turns out that these deformations are quite close to one 
another. 
One must note, however, that only the fluctuating part of the correction for 
the proton and neutron deformation difference can actually influence barrier 
heights calculated with the macroscopic-microscopic approach,
since the average value of this difference can be taken into account in the 
fitting procedure of the parameters of the macroscopic (e.g. liquid drop) 
model. Neglecting the neutron-proton deformation differences in both the 
macroscopic and microscopic parts of the potential energy could then lead 
to an overestimation of spontaneous fission lifetimes by a few orders of 
magnitude in heavy and superheavy nuclei and could also have a non negligible 
effect on the binding energy difference between neighboring nuclei what will 
be reflected in predicted lifetimes for the $\alpha$ or electron-capture 
decays.

The aim of the present investigation is to develop a new term in the liquid
drop type mass formula which will approximate the average variation of the
binding energy when protons and neutrons deform in a different way. It seems to
us that the most suitable approach to this goal is the Extended Thomas--Fermi 
(ETF) approximation \cite{ETF} in connection with the Skyrme energy functional 
\cite{Va72}. 
After a short outline of the model we present results of our calculations for 
a large sample of nuclei and derive a simple approximate expression which 
nicely reproduces the effect of the proton and neutron deformation difference 
in a deformed macroscopic model. 
                                                                  \\[ -2.5ex]
                         
\section{Description of the model}

\subsection{Skyrme density functional and the Thomas--Fermi approximation}

In order to determine nuclear binding energies one often uses effective 
nucleon--nucleon interactions of the Skyrme type which describes quite
accurately nuclear ground--state properties as well as the low-lying collective
excitations \cite{Qu78}. For such an interaction the total energy density
${\cal E}(\vec r)$ is an algebraic function of the neutron and proton densities
$\rho_n$ and $\rho_p$,  of the kinetic energy densities $\tau_n$ and  $\tau_p$
and the spin-orbit densities $\vec J_n$ and $\vec J_p$ \cite{Va72}:

\newpage
\begin{eqnarray}
&& {\cal E}(\vec r)  = \int \left\{ \frac{\hbar^2}{2m} [\tau_n + \tau_p] 
    + \frac{1}{2} t_0 [(1+\frac{1}{2} x_0)\, \rho^2 - (x_0+\frac{1}{2}) 
       (\rho_p^2+\rho_n^2)] \right.
 \nonumber \\
&&  \;\;\;\;\;\;\;\;\;\;\;\; 
    +\frac{1}{4} (t_1+t_2)\rho \, \tau + \frac{1}{8} (t_2-t_1)(\rho_n \, \tau_n
      +\rho_p \, \tau_p) + \frac{1}{16} (t_2 - 3 t_1) \rho \, \nabla^2 \rho 
\label{esky}\\
&&  \;\;\;\;\;\;\;\;\;\;\;\;
      + \frac{1}{32} (3t_1+t_2)(\rho_n \, \nabla^2 \rho_n 
       + \rho_p \, \nabla^2\rho_p) + \frac{1}{4} t_3 \rho \, \rho_n \, \rho_p 
 \nonumber \\      
&&   \left. \;\;\;\;\;\;\;\;\;\;\;
      -\frac{W_0}{2} \left[ \vec{J} \cdot \vec{\nabla} \rho 
                           + \vec{J_n} \cdot \vec{\nabla} \rho_n 
                           + \vec{J_p} \cdot \vec{\nabla} \rho_p \right] 
      +\frac{1}{16}(t_1-t_2)({\vec J}_p^{\,2} + {\vec J}_n^{\,2})      
                                              \right\} d^3r  \;\; , \nonumber
\end{eqnarray}
where non-indexed quantities such as $\rho$ stand for the sum of the 
corresponding quantities for neutrons and protons, as e.g. 
\begin{equation}  
  \rho = \rho_n + \rho_p  
\end{equation}
and where $t_0, t_1, t_2, t_3, x_0$ and $W_0$ are force parameters which can
be adjusted so as to reproduce nuclear ground-state properties (binding
energies, radii, nuclear spectra, etc.). 

In a first approximation the nuclear energy (without Coulomb term) could be
calculated by using the Skyrme  functional (\ref{esky}) and the Thomas-Fermi
(TF) approximation where
\begin{equation}
   \tau_q = \frac{3}{5} (3 \pi)^{2/3} \, \rho_{q}^{5/3} \;\; , 
                \;\;\;\;\;\;\;\;\;\;\; q = \{n,p\} \;\;\; .
\end{equation}
There is no contribution at this (TF) order of the semiclassical expansion 
to the spin-orbit density ${\vec J}_q$ as the spin has no classical analogon. 
These terms are therefore omitted in the following calculations.
                                                                  \\[ -2.5ex]

\subsection{The model with sharp-surface distribution}

If only ellipsoidal deformations are considered, one can describe the surfaces
of nuclei by a very simple parameterization, using a single deformation
parameter $\sigma$. The lengths of the axis of the axially symmetric 
ellipsoides are then given by \cite{My66}~:
\begin{equation}
\left.
\begin{array}{l}
   a = b = R_{0_{q}} \exp\left[-\frac{1}{2}\sigma_{q} \right]\,\,, \\
   c = R_{0_{q}} \exp\left[\sigma_{q} \right] \;\;\; , 
\end{array}  \;\;\;\;\;\;\;\;\;\;\;\;
\right.  q=\{n,p\}
\label{ellips}
\end{equation}
where $R_0^{(n)}$ and $R_0^{(p)}$ are the radii of the neutron and proton 
distributions respectively. Obviously $\sigma > 0$ correspond to prolate and 
$\sigma < 0$ to oblate deformations. Introducing the isospin parameter 
\begin{equation}
   I = \frac{N-Z}{A}
\end{equation}
the following dependence on $A$ and $I$ was obtained in the RMF calculation 
of Ref. \cite{Wa98} 
\begin{equation}
R_{0_{p}}=1.237 \left( 1-0.157 \, I - \frac{0.646}{A} \right) A^{1/3}\,\,
      {\rm fm}\,\,, 
\end{equation}
\begin{equation}
R_{0_{n}}=1.176 \left( 1+0.250 \, I + \frac{2.806}{A} \right) A^{1/3}\,\,
      {\rm fm}\,\,. 
\end{equation}
For ellipsoidal deformations it can be shown that the nuclear surface
is given by 
\begin{eqnarray}
  R_{q}(\theta)=\frac{ac}{\sqrt{a^2\cos^2(\theta)+c^2\sin^2(\theta)}} \;\; . 
\label{surf}
\end{eqnarray}
                                                                   \\[ -2.5ex]

It is obvious that in the case of different deformations and sharp 
distributions of protons and neutrons there will appear regions in which there 
exist only protons or neutrons  and other regions where both types of particles
coexist.
In order to evaluate the nuclear energy in such a case, the Skyrme functional
depending on the nucleon densities, can be separated in three terms depending 
only on the proton density, only on the neutron density and an interaction 
term depending on the densities of both type of particles~:
\begin{equation}
V_q=\frac{1}{4}t_0(1-x_0)\rho_q^2+\frac{1}{8}(t_1+3t_2)\rho_q\,\tau_q
 +\frac{3}{32}(t_2-t_1)\rho_q \, \nabla^2\rho_q \,\,,
\end{equation}
where $q=\{n,p\}$,
\begin{eqnarray}
 V_{p\:n}&=&t_0(1+\frac{x_0}{2})\rho_p\rho_n
          +\frac{1}{4}(t_1+t_2)(\rho_p\tau_n+\rho_n\tau_p) \nonumber\\
         &-&\frac{1}{16}(3t_1-t_2)(\rho_p\nabla^2\rho_n+\rho_n\nabla^2\rho_p)
          +\frac{1}{4}t_3(\rho_n^2\rho_p+\rho_p^2\rho_n) \;\; .
\end{eqnarray}
Finally, ${\cal E}$ can be written as
\begin{equation}
   {\cal E}=\frac{\hbar^2}{2m} (\tau_n + \tau_p) + V_p + V_n + V_{p\:n} \;\; .
\end{equation}
The total nuclear energy $E_{nuc}(N,Z)$ is then given as the volume integral 
of ${\cal E}(\vec r)$.
\begin{equation}
   E_{nuc}(N,Z)=\int {\cal E}(\vec r) \, d^3r \;\; .
\end{equation}
In the case of a sharp surface distribution the changes of the nuclear part of 
the energy as a function of deformation can be just expressed as the product 
of the volume $\Omega_{p\,n}$ where protons and neutrons coexist and the term 
of the energy density connected with the interaction of these two types of 
particles~:
\begin{eqnarray}
\Delta E_{nuc}(\Delta\sigma)=V_{p\:n}\,\Omega_{p\,n} \;\; ,
\end{eqnarray}
where
\begin{eqnarray}
   \Omega_{p\,n} & = & \frac{4}{3}\pi (R_{0_{p}})^3
     \left\{\left[1-\frac{(\frac{R_{0_{p}}}{R_{0_{n}}})^2
     \exp(\Delta\sigma)-1}{\exp(3\Delta\sigma)-1}\right]^{3/2}-1\right\} \\
   & - & \frac{4}{3}\pi
     (R_{0_{n}})^3\left\{\left[1-\frac{(\frac{R_{0_{n}}}{R_{0_{p}}})^2
     \exp(-\Delta\sigma)-1}{1-\exp(-3\Delta\sigma)}\right]^{3/2}-1\right\}  
\nonumber
\end{eqnarray}
with $\Delta\sigma=\sigma_n-\sigma_p$.

One should notice that for this type of nucleon densities, the terms of 
${\cal E}(\vec r)$ with $\Delta \rho$ vanish, so that the nuclear energy 
doesn't change during the deformation process as long as the proton and 
neutron distribution don't cross each other, since the volumes between the 
sharp surfaces is kept constant. After the surfaces intersect the energy 
rapidly decrease because there the region in which protons and neutrons 
coexist $\Omega_{p\:n}$ changes its volume as a function of the deformation. 
This fact leads to the changes of the term $V_{p\:n}$,
while terms $V_{p}$ and $V_{n}$ are constant. A typical behavior of the
binding energy with growing proton--neutron deformation difference is
presented in Fig. 1. The model with the uniform density distribution is, of 
course, too rough to make a realistic estimates but nevertheless it should 
give some idea about the main effect.
                                                                   \\[ -2.5ex]

\subsection{The model with the Fermi-type distribution.}

More accurate estimates of the change of the binding energy when the protons
and neutrons deform differently can be made with diffuse density profiles. 
In the following we have chosen for our analysis the density distributions 
in the form of Fermi functions
\begin{equation}
   \rho_q(r,\theta) = \frac{\rho_{0_{q}}}
                {1+\exp(\frac{r-R_{q}(\theta)}{a_q(\theta)})} \;\; ,
\label{fermi}
\end{equation}
where $\rho_{0_{q}}$ are the saturation density parameters obtained from the 
normalization conditions~:
\begin{equation}
\int\rho_p(r)d^3r = Z \;\; , \;\;\;\;\;\;\;\;\;\;\;\;
\int\rho_n(r)d^3r = N 
\end{equation}
and where $R_{q}(\theta)$ is given by Eq.~(\ref{surf}). The surface width 
parameters $a_q$, are $\theta$ dependent and equal to
\begin{equation}
\!\!\!\!\!\!\!\!\!\!\!\!\!\!\!\!\!\!a_{q}(\theta)=
 \sqrt{\sin^4(\theta)+\cos^4(\theta)+\sin^2(\theta)\cos^2(
\theta)[\exp(3\sigma_{q})+\exp(-3\sigma_{q})]\;} \;\, a_{0_{q}}
\end{equation}
a definition which guarantees that the surface diffuseness is constant along 
the direction perpendicular to the surfaces of the ellipsoids. The parameters 
$\{ R_{0_{n}}, a_{0_{n}}, R_{0_{p}}, a_{0_{p}}\}$ characterizing the spherical 
density distribution of neutrons and protons are obtained by a minimization 
procedure of the energy functional (\ref{esky}) within the second order ETF 
approximation \cite{ETF} and using the Skyrme SIII force \cite{BFN75}.
                                                                   \\[ -2.5ex]
 
\section{Results of the fits}

The calculation were performed for 28 even-even nuclei involving several 
isotopic chains from Ni to Th.
We suppose that the deformation of these nuclei can be characterized by a
single {\it global} deformation parameter $\alpha$ plus a quadrupole type 
deformation parameter $\Delta\tilde\beta = \tilde\beta_n - \tilde\beta_p$ 
measuring the difference in the proton versus neutron deformation. 
The global deformation of the nucleon distribution can be e.g. defined through 
the parameter $\alpha$ introduced by Myers and \'Swi\c{a}tecki \cite{My66} 
\begin{equation}
\alpha^2= 2\pi\,\int\limits_0^\pi d\theta\left[\frac{R(\theta)-R_{00}}{R_{00}}
          \right]^2  \,\,,
\end{equation}
where $R(\theta)$ is the half-density radius of the nucleon distribution of 
the deformed nucleus and $R_{00}$ the radius of the corresponding spherical 
distribution. For an ellipsoidal deformation as characterized by Eq. 
(\ref{ellips}) $\alpha$ can be easily expressed in terms of the parameter 
$\sigma$ and one obtains
\begin{equation}
  \alpha^2 = \sigma^2 [1-\frac{1}{7}\sigma + {\cal O}(\sigma^2)] \;\; .
\end{equation}
In order to be able to express the total energy as a function of the 
difference $\Delta\tilde\beta$ between the proton versus neutron deformation 
we need to find a way to express this difference in a way which is independent
of the specific choice of the deformation parameters like e.g. in Eq. 
(\ref{ellips}). We have chosen to define parameters $\tilde\beta_q$ as~:
\begin{equation}
   \tilde\beta_q = \frac{Q_{20}^{(q)}}{\tilde Q_{00}^{(q)}} \;\; ,
\end{equation}
where the $Q_{0 0}^{(q)}$ are the monopole and $Q_{2 0}^{(q)}$ the 
quadrupole moments of the proton and neutron distribution with~: 
\begin{equation}
   Q_{\lambda 0}^{(q)} = \int \rho_q (\vec r) \, r^2 \, P_\lambda \, d^3r \,\,,
\end{equation}
and where 
\begin{eqnarray}
Q_{00}^{tot}=Q_{00}^{(n)} + Q_{00}^{(p)} \,\,,                 \nonumber \\ 
\tilde Q_{00}^{(p)}=Q_{00}^{tot}\cdot\frac{Z}{A} \,\,,           \\
\tilde Q_{00}^{(n)}=Q_{00}^{tot}\cdot\frac{N}{A} \,\,. \nonumber 
\end{eqnarray}
are the so-called average monopole moments for protons and neutrons 
respectively.

We have found that contrary to the rapid change (Fig. 1) of the binding 
energy in case of the sharp-edge density distribution the nuclear part of
energy varies almost parabolically with the proton and neutron deformation 
difference $\Delta\tilde\beta$ when the smooth density profile (\ref{fermi}) 
is assumed for protons and neutrons. 

The binding energy can then be parametrized in the following form 
\begin{equation}
   B(A,I,\alpha,\tilde\beta_n,\tilde\beta_p) = B_{avr}(A,I,\alpha) 
                  + a(A,I,\alpha) \cdot (\tilde\beta_n - \tilde\beta_p)^2 \;\; ,
\label{B}\end{equation}
where $B_{avr}$ is the part of the energy that is generated for {\it equal}
deformations for protons and neutrons and which can be obtained e.g. form 
the standard liquid drop or droplet model. 
To determine the stiffness parameter $a$ of the parabolic form in 
$\Delta \tilde\beta$ we proceed in the following way~:
\\
For  various nuclei with different values of the mass parameter $A$ and isospin 
parameter $I$ and for $\alpha \neq 0$ one finds that this parameter 
$a(A,I,\alpha)$ can be expressed in the form 
\begin{equation}
   a(A,I,\alpha) = a_0(A,I) \; [1 - c_2\,\alpha(1+c_3\,I)] \;\; ,
\label{A}
\end{equation}
where the deformation independent coefficient $a_0(A,I)$ can be parametrized as  
\begin{equation}
   a_0(A,I)=c_1  \left( A^{1/3} \right)^n \;\; .
\label{A0}
\end{equation}
The values found by our fitting procedure for the parameters $c_1$, $c_2$, $c_3$
and $n$ are the following
$$
   c_1 = 1.21 \, {\rm MeV} \;\; , \;\;\;\;\;\;  c_2 = 3.0 \;\; , \;\;\;\;\;\; 
   c_3 = 3.0 \;\; , \;\;\;\;\;\;   n = 4   \;\; .
$$
The stiffness parameter $a(A,I,\alpha)$ evaluated for all considered isotopic 
chains is compared on the l.h.s. of Fig. 2 at $\alpha \!=\! 0$ deformation 
with the parameter $a_0(A,I)$, Eq. (\ref{A0}). The $\alpha$ deformation 
dependence of the parameter $a(A,I,\alpha)$ is illustrated on the r.h.s. of 
Fig. 2.
We have also compared the above estimate of $a(A,I,\alpha)$ with the
results obtained using two other Skyrme interactions. We have found
that the estimates obtained with the Skyrme SII and SVII forces are 
identical to those of presented above for SIII within a 10\%. 

The quality of the fit of the binding energy as function of the deformation 
difference is demonstrated on Fig.~3 for three nuclei ($^{98}$Zr, $^{146}$Nd 
and $^{208}$Pb) and for three different values of the global deformation 
parameter $\alpha$ and compared with the corresponding ETF results. 

In order to test the predictive power of equation (\ref{B}) we have
performed an additional calculation for three dysprosium isotopes (neutron 
deficient, $\beta$-stable and neutron rich) and compare in Fig. 4 their ETF 
energies (crosses) with the approximation given by (\ref{B}) (solid line)  
and this for three different deformation. As can be seen in Fig. 4 the 
agreement of both results is rather satisfactory in all cases, except for the 
very deformed neutron deficient $^{150}$Dy isotope.

When $\alpha\neq 0$ and $(\tilde\beta_n-\tilde\beta_p)=0$, the energy is
reproduced by the standard liquid drop model or any other macroscopic model.
                                                                   \\[ -2.5ex]
\section{Discussion and conclusions}
 
We have found a very simple procedure to describe the influence on the total
nuclear energy arising from the fact that neutron and protons might deform 
differently. The additional term in the nuclear mass formula which we obtain 
is proportional to the square of the deformation difference of of both 
distributions and increases as $A^{4/3}$. 

This suggest that it might be useful to generalize the currently used 
macroscopic-microscopic approaches in order to allow protons and neutrons to
have different multipole deformations. One could start from a Strutinsky type 
prescription :
\begin{equation}
 E_{\rm Strut}(\{\beta^{(p)}_\lambda\}, \{\beta^{(n)}_\lambda\}) =
 E_{\rm macr}(\{\beta^{(p)}_\lambda\}, \{\beta^{(n)}_\lambda\}) +
\delta E^{(p)}_{\rm micr} (\{\beta^{(p)}_\lambda\}) 
   + \delta E^{(n)}_{\rm micr} (\{\beta^{(n)}_\lambda\}) \;\; ,
\end{equation}  
where both the macroscopic and and microscopic components depend on proton and 
neutron deformations. Then, performing a minimization with respect to e.g., 
the neutron multipole deformations while keeping proton deformations as 
independent variables (or the other way around), should give a more realistic 
description of potential energy surfaces. 
Details of such a method as, for instance, the way to generalize
macroscopic models in order to include different deformations for protons and
neutrons, are left for a future study. 

It was already shown in Ref.~\cite{Be00} that taking into account the proton 
versus neutron deformation difference could shift the position of the ground 
state and saddle point  in the potential energy surface by as much as 0.5 MeV.
Such an effect could improve the predictions of nuclear masses as given 
by macroscopic--microscopic type of models \cite{Mo95} and change the 
predicted barrier heights. In consequence the effect discussed in the present 
paper could change the theoretical predictions of $\alpha$-decay and
spontaneous-fission life times \cite{Sm95,St99}.

\bigskip\bigskip    
\noindent
{\bf Acknowledgements}
\bigskip                     
 
Two of us (A.D. and K.P.) are very grateful for the nice hospitality extended 
to them by the Nuclear Theory Group of the IReS in Strasbourg.

\newpage
 
\centerline{\large {\bf Figure captions}}
 
\bigskip
\begin{enumerate}

\item  Typical behavior of the binding energy with growing neutron--proton
       deformation difference for the nucleus $^{208}$Pb {and the sharp 
       proton and neutron density distributions.} 
       
\item  The mass number (l.h.s. figure) and the deformation (r.h.s. figure)
       dependence (crosses) of the stiffness parameter $a(A,I,\alpha)$ 
       (Eq. \ref{B})) and its approximation by the formula (\ref{A}) (solid 
       lines).

\item  The change of the binding energies due to different neutron--proton
       deformations as obtained in the ETF approach is compared with the 
       approximate expression of Eq. (\ref{A}).

\item  The approximation (\ref{B}) of the binding energy (solid line) made
       for three Dy isotopes and at three deformation points is compared with 
       the ETF results (crosses) as function of the neutron-proton 
       deformation difference.

\end{enumerate}

\pagestyle{empty}
\newpage
\begin{figure}
\epsfxsize=100mm \epsfbox{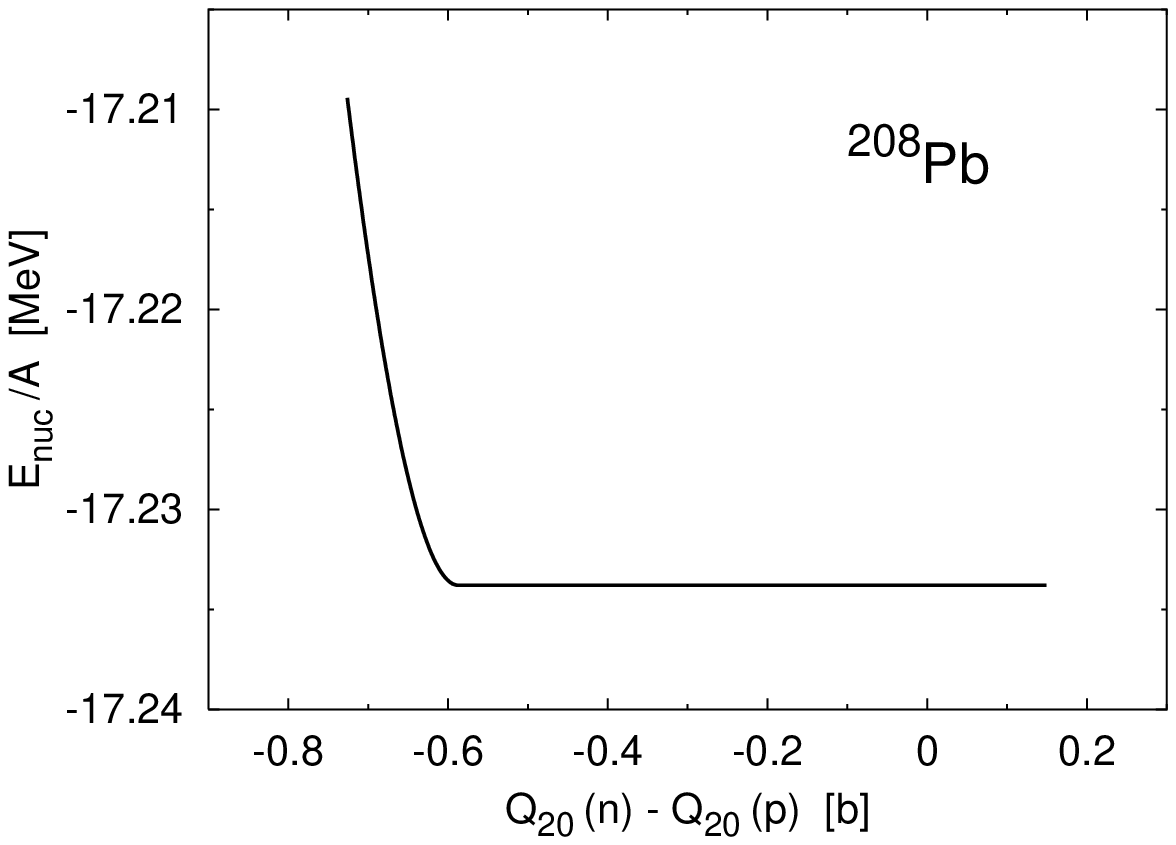}
\caption{ }
\end{figure}

\pagebreak[5]
\begin{figure}
\epsfxsize=140mm \epsfbox{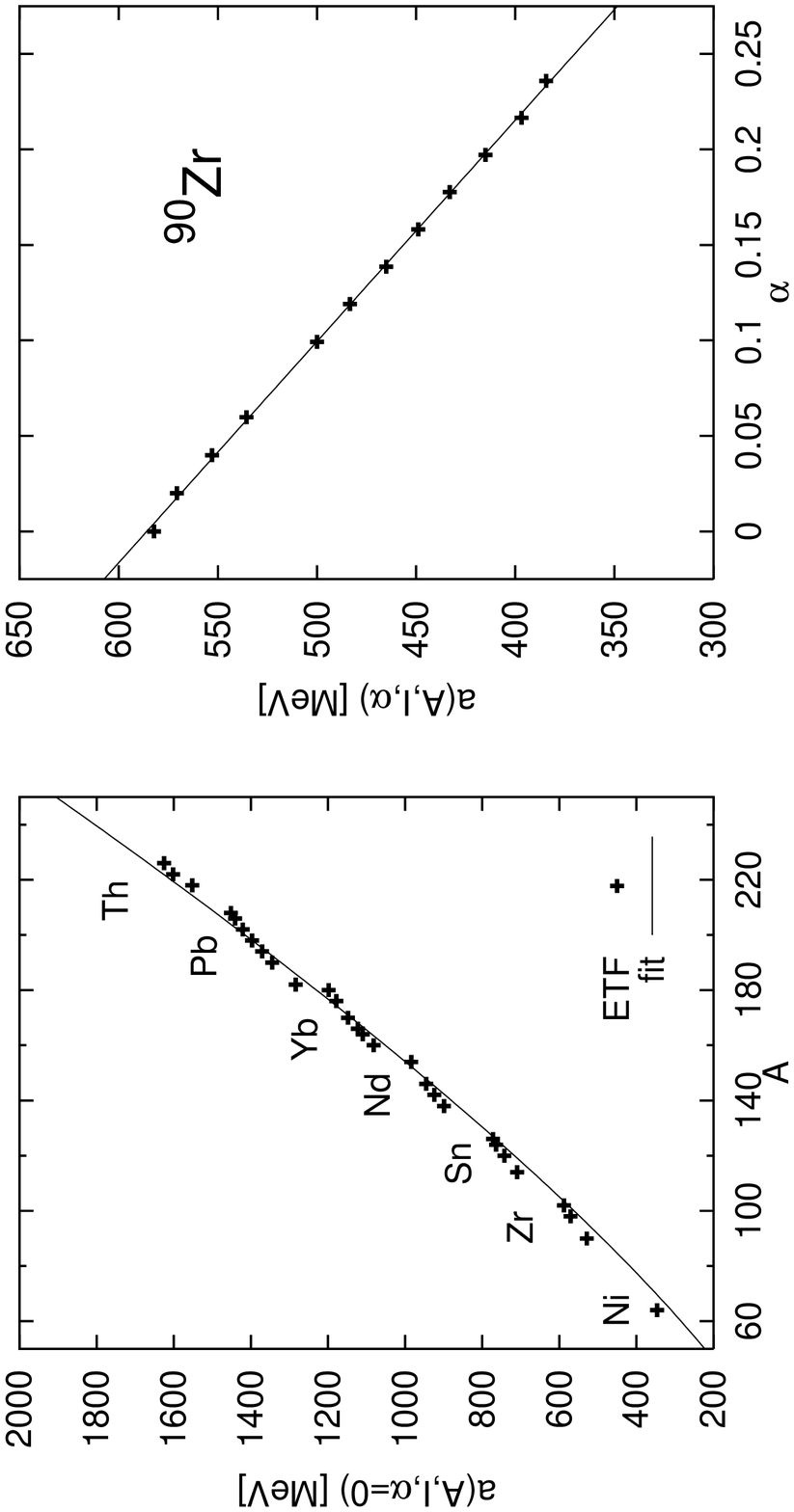}
\caption{ }
\end{figure}
 
\pagebreak[5]
\begin{figure}
\epsfxsize=140mm \epsfbox{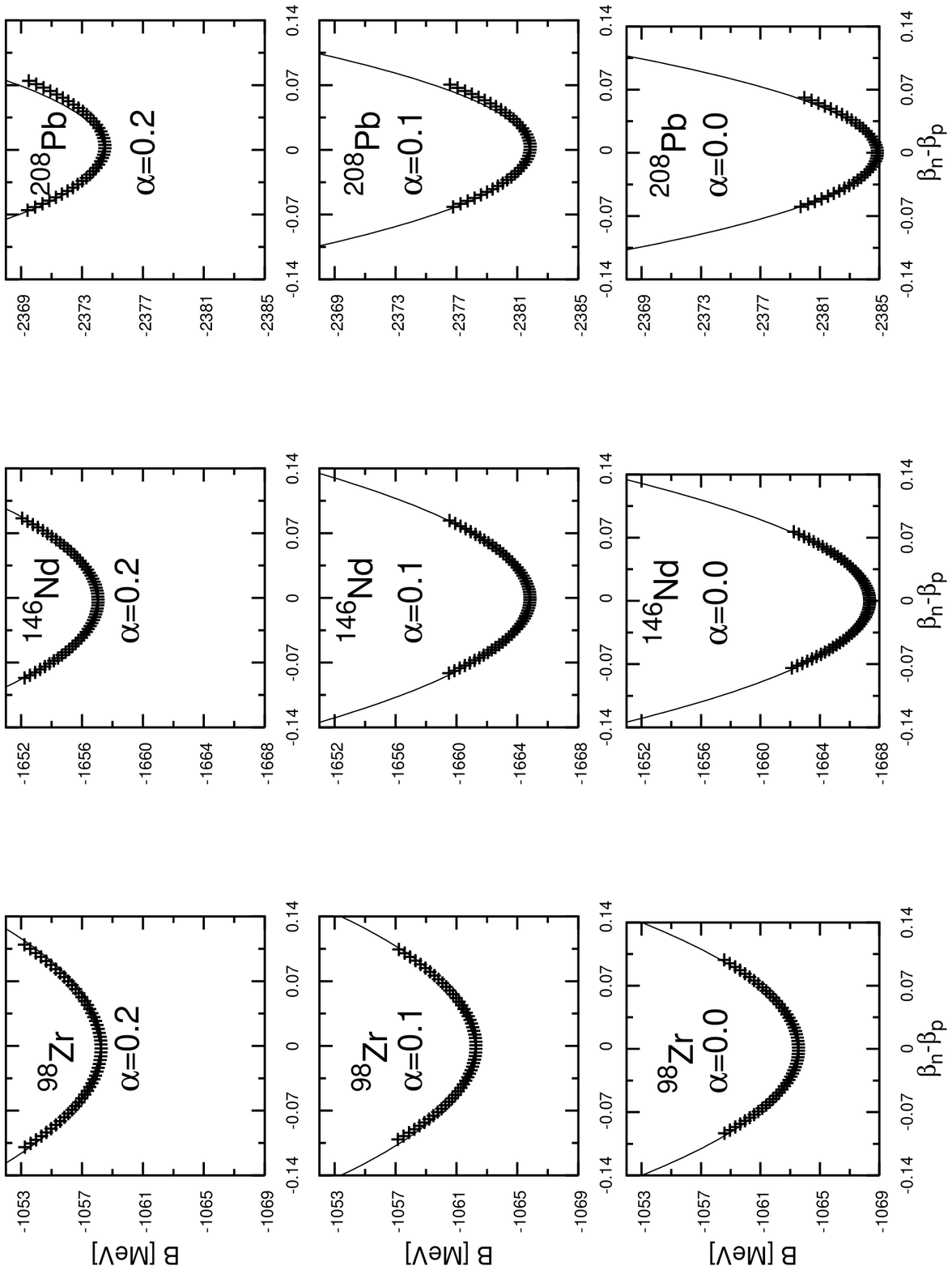}
\caption{ }
\end{figure} 

\pagebreak[5]
\begin{figure}
\epsfxsize=140mm \epsfbox{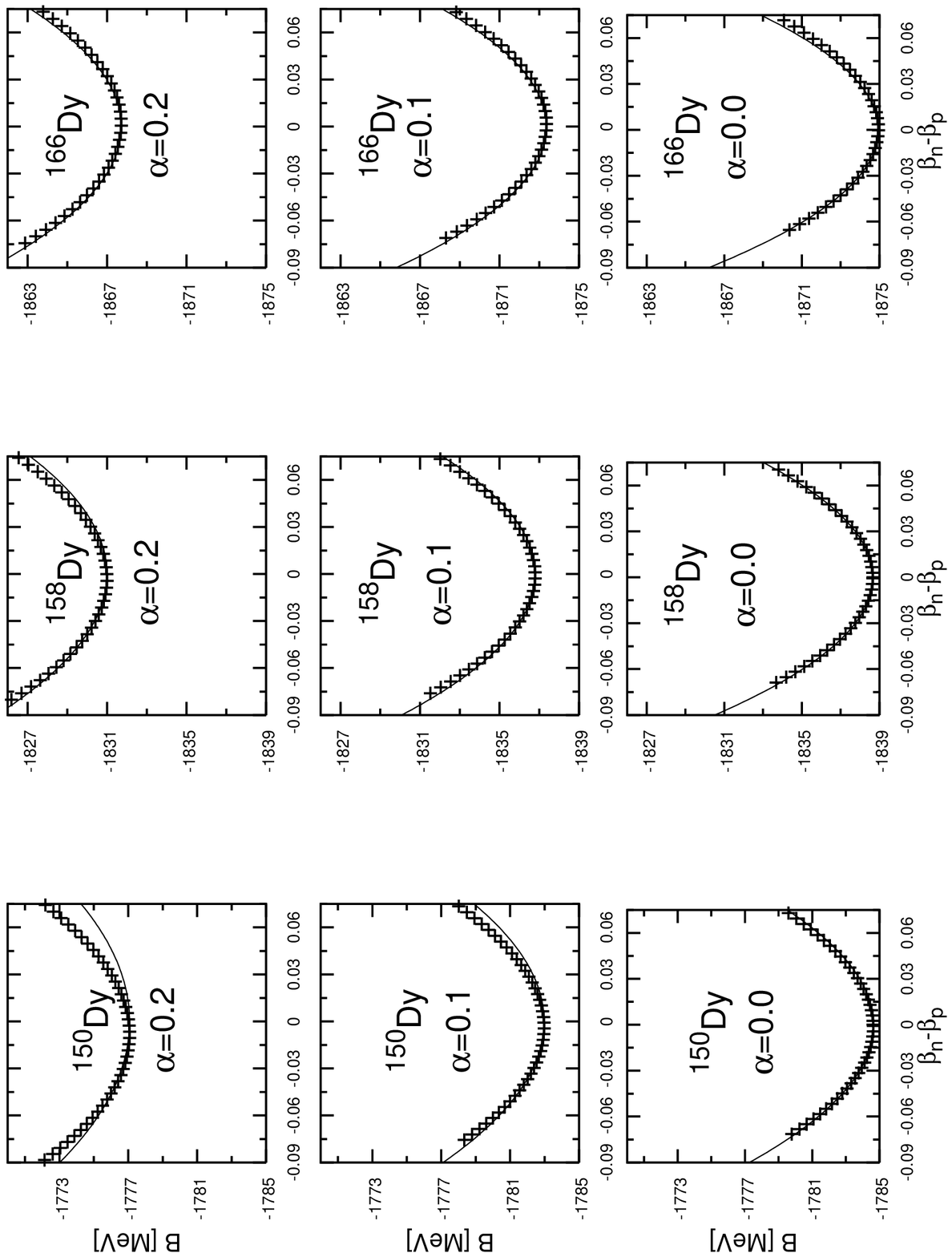}
\caption{ }
\end{figure} 

\end{document}